\documentstyle{article}

\title{Physics of a microsystem starting from non-equilibrium
quantum statistical mechanics}

\author{L.~Lanz${}^1$, O.~Melsheimer${}^2$ and
B.~Vacchini${}^{1}$
\\
${}^1$
Dipartimento di Fisica
dell'Universit\`a di Milano and INFN,
Sezione di Milano,
\\
Via Celoria 16, I--20133, Milan, Italy
\\
${}^2$
Fachbereich Physik, Philipps-Universit\"at,
Renthof 7, D--35032, Marburg, Germany
\\
(e-mail:
lanz@mi.infn.it;
melsheim@mailer.uni-marburg.de;
vacchini@mi.infn.it)
\\[2ex]
}

\begin{document}
\maketitle
\begin{abstract}
In this paper we address the problem to give a concrete support to the
idea, originally stemming from Niels Bohr, that quantum mechanics must be
rooted inside the physics of macroscopic systems. It is shown that,
starting from the formalism of the non-equilibrium statistical operator,
which is now a consolidated part of quantum statistical
mechanics, particular correlations between two isolated systems can be
singled out and interpreted as microsystems. In this way also a new
framework is established in which questions of decoherence can be
naturally addressed.

\end{abstract}

\section{Introduction}
The motivation of this paper is twofold: on the one side the question
about foundations of quantum mechanics still catches the
interest~\cite{Bonifacio-Decoherence} of many scientists, mainly due
to important 
improvements in experimental physics and to the increasing relevance
of the concept of entanglement in the realm of quantum computation; on
the other side there is relevant progress in the theory of
non-equilibrium statistical mechanics~\cite{Roepke-Zubarev}. That a deep
connection between these two apparently unrelated subjects exists
should be clear if one thinks about the way Bohr discussed foundations
of quantum mechanics, facing Einstein's objections against quantum
mechanics at the beginning of all questions about reality of
microsystems. Let us recall that Bohr's main concern was the very
rooting of microsystems inside objectively given macroscopic reality.
This point of view has been pursued further by Ludwig~\cite{Ludwig},
who starting from a purely phenomenological macroscopic setting was
able to derive and extend the formalism of quantum mechanics, giving
an insight into what should be understood as reality of microsystems.
In this connection also a more profound theory of measurement, with
respect to which textbook measurement theory appears as a very naive
idealization, has grown up and has found many applications, typically
in quantum optics, by the work of many
researchers~\cite{misura}. A very
systematic and comprehensive account of all this is given by Holevo's
work~\cite{Holevo}, who contributed to all relevant steps. 

Despite the formal achievements of this more realistic axiomatic
structure, the basic necessity of this approach and its real
effectiveness to cope successfully difficulties in the foundations of
quantum mechanics is still not in tune with the general idea that only
the physics of elementary particles should have a fundamental role,
any compelling resorting to macrophysics being only an annoying
bypass. Our attitude instead is that the consideration of macroscopic
physics gives a strong motivation in favor of the first standpoint: in
fact any real experiment is described and realized in terms of
macrosystems. The basic tool for this is quantum field theory for
confined systems: the extraction of the concept of particles and of
their local interactions could be a subsequent concern, to be perhaps
performed using the typical procedures of a thermodynamic limit.

Obviously our attempt can only start if a working theory of
non-equilibrium macrosystems exists. In this connection the situation
has improved relying on the concept of relevant variables and on the
method of the non-equilibrium statistical
operator~\cite{Roepke-Zubarev}.  In previous papers some improvement
in the foundations of this concept was proposed insisting on the ideas
of confinement, isolation and preparation of a
macrosystem~\cite{gargnano99-torun99} during some initial time interval.
A rather general and tractable situation, that we called ``simple
dynamics'', arises if the evolution of the local relevant variables is
driven by two-point Kubo correlation functions involving relevant
variables and their currents at points separated by a short time
interval: in other words memory contributions are assumed to be
restricted to short time intervals. On the contrary if two isolated
macrosystems exchange a particle during the preparation stage a
possibly long living correlation between the systems arises if this
particle presents a coherent independent dynamics for a suitable time
interval: in the present paper just this situation is discussed. 

The treatment seems to us interesting for two reasons. On one side the
microphysical description can be read off from the macroscopic
dynamics of the system: the initial state of the microsystem can be
related to a macroscopic source, the time evolution of the microsystem
is part of the dynamics of the statistical operator, the measurement
of an observable for the microsystem is related to the values of
relevant variables of the macrosystem; obviously all this is treated
only in a very schematic way. On the other side one gains a way to
face the case of long lived correlations using the dynamics of the
microsystem, thus providing an example in which the difficult problem
of a dynamics with memory can be treated. A final remark is now in
order; in our description a microsystem unavoidably appears having a
macrosystem in the background: then its time evolution is necessarily
affected by the decoherence phenomenon.  Our treatment gives a natural
reason to decoherence, so that is does not appear as an additional
feature of quantum mechanics or as a motivation to modify the theory.

The paper is organized as follows: in Sect.~2 we introduce the
statistical operator describing a system composed of two parts which
are isolated from each other after the preparation procedure, but keep
record of a correlation arising during the preparation time; the
correction of the dynamics due to this correlation is worked out in
Sect.~3 and interpreted as a microsystem produced in system 2,
propagating and detected in system 1. Problems related to the dynamics
of the microsystem inside a macrosystem are outlined in Sect.~4.

\section{Separation of two isolated macrosystems and initial correlations}
We rely in the following on a general theory of an isolated
macroscopic system inside quantum field theory that has been proposed
in previous papers~\cite{gargnano99-torun99} and is very close to statistical
mechanics formulated in terms of the non-equilibrium statistical
operator by Zubarev and more recently by Morozov and
Roepke~\cite{Roepke-Zubarev}, differing only in the introduction of a
``preparation procedure'' which extends over a finite time interval and
refers to a confined isolated system, thus avoiding in principle the
thermodynamic limit at the level of foundations of macroscopic
physics. According to this theory a set of relevant variables is selected,
slow enough on a suitable time scale, for which the very concept of
isolation can make sense, built in terms of suitable field
densities. Let us take $[T,{t_0}]$ as preparation time interval, the
statistical operator at time $t_0$ is given by
        \begin{eqnarray}
        \label{1}
        {\hat \varrho}_{t_0}
        \!\!
        \!\!
        &=&
        \!\!
        \!\!
        \exp
        \left \{
        -\sum_j
        \int d \textbf{x} \,
        \zeta_j (\textbf{x},{t_0})
        {\hat A}_j(\textbf{x})
        +
        \sum_{j \alpha}
        \int d \textbf{x}  \,
        \gamma_{j\alpha} (\textbf{x})
        \int_T^{t_0}  dt' \,
        {\hat A}_j(\textbf{x},-(t_0 - t'))
        h_{j\alpha}(t')
        \right.
        \nonumber
        \\
        &&
        \!\!
        \!\!
        \hphantom{
        \exp
        \left \{
        \right.
        }
        +
        \sum_{j \alpha}
        \int d \textbf{x} \,
        {\mbox{{\boldmath $\gamma$}}}_{j\alpha} (\textbf{x})
        \cdot
        \int_T^{t_0} dt' \,
        {\hat {\bf J}}_j(\textbf{x},-(t_0 - t'))
        h_{j\alpha}(t')
        \nonumber
        \\
        &&
        \!\!
        \!\!
        \hphantom{
        \exp
        \left \{
        \right.
        }
        \left.
        -\sum_j
        \int d \textbf{x} \,
        \zeta_j (\textbf{x},T)
        {\hat A}_j(\textbf{x},-(t_0 -T))
        \right \}
        ,
        \end{eqnarray}
with $\zeta_j (\textbf{x},{t_0})$ the classical state
parameters such that
\begin{displaymath}
        {\mbox{{\rm Tr}}} \,
        \left(
        {\hat A}_j(\textbf{x})
        {\hat \varrho}_{t_0}
        \right)
        =
        {\mbox{{\rm Tr}}} \,
        \left(
        {\hat A}_j(\textbf{x})
        {{\hat w}_{\zeta{({t_0})}}}
        \right)
        ,  
\end{displaymath}
where ${{\hat w}_{\zeta{(t)}}}$ is the generalized Gibbs state
related to the state parameters ${\zeta_j{(t)}}$
\begin{displaymath}
          {\hat w}_{\zeta(t)}
        =
        {  
        \exp
        \left \{
        {-{
        \sum_j
        \int d \textbf{x} \,
        {\zeta}_j(\textbf{x},t)
        {\hat A}_j(\textbf{x})
        }}  
        \right \}
        \over 
        {\mbox{{\rm Tr}}}
        \left(        \exp
        \left \{
        {-{
        \sum_j
        \int d \textbf{x} \,
        {\zeta}_j(\textbf{x},t)
        {\hat A}_j(\textbf{x})
        }}  
        \right \}\right)
        }
.
\end{displaymath}
${\hat A}_j(\textbf{x})$ are the densities of the
relevant variables, associated to the currents ${\hat {\bf
    J}}_j(\textbf{x})$ and ${\mbox{{\boldmath
      $\gamma$}}}_{j\alpha}$, $ h_{j\alpha}$ are the parameters
describing the preparation procedure. We assume for simplicity
that all the field variables are built with only one underlying quantum
Schr\"odinger field ${\hat \psi}({\mbox{\bf x}})$, satisfying
\begin{displaymath}
        \left[  
        {\hat \psi}({\mbox{\bf x}}),
        {\hat
        \psi}^{\scriptscriptstyle\dagger}({\mbox{\bf
        x}}')
        \right]_\pm  
        =  
        \delta^3 ({\mbox{\bf x}}-{\mbox{\bf x}}').
      \end{displaymath}
In this context one can very easily formalize the
idea that a macroscopic system is separated into two non interacting
parts 1 and 2 corresponding to two non overlapping regions $\omega_1$,
$\omega_2$. For times $t\geq t_0$ let us introduce two complete sets
of normal modes for both regions: 
$\{ u_n^{\scriptscriptstyle (1)} (\textbf{x}),
\textbf{x}\in\omega_1\}$ and $\{ u_n^{\scriptscriptstyle (2)} 
(\textbf{x}), \textbf{x}\in\omega_2\}$, determined by
\begin{displaymath}
          -{
        \hbar^2  
        \over  
        2m
        }  
        \Delta_2  
        u_n^{\scriptscriptstyle (1)} (\textbf{x}) +V (\textbf{x})
        u_n^{\scriptscriptstyle (1)} (\textbf{x})=
        W^{\scriptscriptstyle (1)}_n u_n^{\scriptscriptstyle (1)}
        (\textbf{x}) 
        \quad \textbf{x}\in \omega_1,
        \qquad  
        u_n^{\scriptscriptstyle (1)} (\textbf{x})=0 \quad\textbf{x} \in
        \partial\omega_1
\end{displaymath}
and similarly for $u_n^{\scriptscriptstyle (2)} (\textbf{x})$. Finally
we  build the two field operators 
\begin{equation}
\label{3}
        {\hat \psi}^{(1,2)}({\mbox{\bf x}})
        =
        \sum_n
        {\hat a}^{(1,2)}_{ n}
        u_n^{(1,2)}({\mbox{\bf x}})
,
\quad
\textrm{where}
\quad
        [  
        {\hat a}^{(1,2)}_{ n},
        {\hat a}^{(1,2){\scriptscriptstyle\dagger}}_{ n'}
        ]_\pm  
        =  
        \delta_{n,n'}
\end{equation}
and the creation and destruction operators referring to the two
regions $\omega_1$ and $\omega_2$ commute (or anticommute).
Relevant variables ${\hat A}_j(\textbf{x})$ which are
generally given for the non confined fields in terms of ${\hat
  \psi}({\mbox{\bf x}})$, ${\hat
  \psi}^{\scriptscriptstyle \dagger}({\mbox{\bf x}})$ are now
expressed in terms of ${\hat \psi}^{(1,2)}({\mbox{\bf x}})$, ${\hat
  \psi}^{(1,2){\scriptscriptstyle \dagger}}({\mbox{\bf x}})$
according to $\textbf{x}\in\omega_1$ or  $\textbf{x}\in\omega_2$. This
seems to be the most straightforward way of partitioning the system
into two non interacting parts and to represent the physical setup that
is necessary for this: also different boundary conditions are
conceivable, our choice represents the perfectly reflecting walls. In
our point of view, putting confined systems in the foreground,
boundary conditions have a very important role and unconfined field
theory is only the starting point to introduce concrete
descriptions. Since the Hamiltonian of a system extending on a region
$\omega$ is given by
        $
        {\hat H}=
        \int_{\omega} d^3\!
        {\mbox{\bf x}}
        \,
        {\hat e}({\mbox{\bf x}})
        $
one has ${\hat H}={\hat H}^{\scriptscriptstyle (1)}
+{\hat H}^{\scriptscriptstyle (2)}$ where ${\hat
  H}^{\scriptscriptstyle (1)}$, ${\hat H}^{\scriptscriptstyle (2)}$ is
expressed 
only in terms of ${\hat a}^{\scriptscriptstyle (1)}_{ n}$, ${\hat
  a}^{\scriptscriptstyle (1){\scriptscriptstyle\dagger}}_{ n}$ and
${\hat a}^{\scriptscriptstyle (2)}_{ n}$, ${\hat 
  a}^{\scriptscriptstyle (2){\scriptscriptstyle\dagger}}_{ n}$ 
respectively, so
that $[{\hat H}^{\scriptscriptstyle (1)},{\hat H}^{\scriptscriptstyle
  (2)}]=0$; furthermore, since ${\hat a}^{\scriptscriptstyle (1)}_{
  n}$, ${\hat 
  a}^{\scriptscriptstyle (1){\scriptscriptstyle\dagger}}_{ n}$ and
${\hat a}^{\scriptscriptstyle (2)}_{ n}$, ${\hat 
  a}^{\scriptscriptstyle (2){\scriptscriptstyle\dagger}}_{ n}$ commute
(or anticommute) all the relevant variables $ {\hat
  A}_j(\textbf{x})$ and $
        {\hat A}_j(\textbf{x},\tau)
        =
        e^{+{{
        i
        \over
         \hbar
        }}{\hat H}\tau}
        {\hat A}_j(\textbf{x})
        e^{-{{
        i
        \over
         \hbar
        }}{\hat H}\tau}
$ have a tensor product structure:
\begin{displaymath}
  \int_{\omega_1 \cup\omega_2} d^3\!{\mbox{\bf x}}\,
  {\hat A}_j(\textbf{x},\tau)
\beta_j (\textbf{x})=
  \int_{\omega_1} d^3\!{\mbox{\bf x}}\,
  {\hat A}_j^{\scriptscriptstyle (1)} (\textbf{x},\tau)
\beta_j^{\scriptscriptstyle (1)} (\textbf{x})
\otimes
{\hat 1}^{\scriptscriptstyle (2)}
+
{\hat 1}^{\scriptscriptstyle (1)}
\otimes
  \int_{\omega_2} d^3\!{\mbox{\bf x}}\,
  {\hat A}_j^{\scriptscriptstyle (2)} (\textbf{x},\tau)
\beta_j^{\scriptscriptstyle (2)} (\textbf{x})
\end{displaymath}
with $\beta_j^{\scriptscriptstyle (1)}
(\textbf{x})=\beta_j({\textbf{x}})$, $\textbf{x}\in\omega_1$,
$\beta_j^{\scriptscriptstyle (2)} (\textbf{x})=\beta_j({\textbf{x}})$,
$\textbf{x}\in\omega_2$.  If we assume that such a separation between
system 1 and system 2 also occurs during the preparation time interval
$[T,{t_0}]$, i.e., the operators ${\hat A}_j(\textbf{x},-(t_0 - t'))$
and ${\hat {\bf J}}_j(\textbf{x},-(t_0 - t'))$, $t'\in [T,{t_0}]$
which appear in (\ref{1}) also have the structure (\ref{3}), one has
the factorized structure
\begin{displaymath}
{\hat \varrho}_{t_0}  =
{\hat \varrho}^{\scriptscriptstyle (1)}_{t_0}\otimes {\hat
  \varrho}^{\scriptscriptstyle (2)}_{t_0} 
\end{displaymath}
with ${\hat \varrho}^{\scriptscriptstyle (1)}_{t_0}$ (${\hat
  \varrho}^{\scriptscriptstyle (2)}_{t_0}$) having the structure
(\ref{1}) with region $\omega$ replaced by $\omega_1$ ($\omega_2$).
Obviously if this situation has been obtained by a previous
preparation procedure implementing the separation, one has to assume
that correlations between 1 and 2 are already decayed at time $T$.
This kind of assumption is anyway necessary to start any description
of isolated systems. Now we are interested in a more sophisticated
situation: during the preparation time $[T,{t_0}]$ some very
particular interaction arose, which eventually we will describe as due
to a microsystem coming from system 2 and effecting system 1 at times
$t>t_0$. In this way system 2 becomes a source and system 1 a detector
for the microsystem. The difference in the procedure of isolating
system 1 for times $t>t_0$ in the two cases: already isolated from
part 2 during $[T,{t_0}]$ or interacting with part 2 during $[T,{t_0}]$,
has relevant consequences on the dynamics of the system; in fact in
the second case the time scale of the processes happening inside the
detector should be taken into account. From a formal point of view in
the new situation the time dependence of the operators in (\ref{1})
should be ruled by an effective Hamiltonian, which no longer has the
structure (\ref{3}). Now our aim is not to give a detailed description
of how the source works, but to indicate key points characterizing
this kind of coupling and correlation between the two systems.

At time $t_0$ the statistical operator ${\hat \varrho}_{t_0}$ should
be represented as:
\begin{displaymath}
  {\hat \varrho}_{t_0}=
{
\exp {\{- [{\hat S}^{\scriptscriptstyle (1)}_{t_0}+{\hat
    S}^{\scriptscriptstyle (2)}_{t_0}+{\hat C}^{\scriptscriptstyle
    (12)}_{t_0}] \}} 
\over
\textrm{Tr}\left[
\exp {\{- [{\hat S}^{\scriptscriptstyle (1)}_{t_0}+{\hat
    S}^{\scriptscriptstyle (2)}_{t_0}+{\hat C}^{\scriptscriptstyle
    (12)}_{t_0}] \}} 
\right]
}
\end{displaymath}
where ${\hat C}^{\scriptscriptstyle (12)}_{t_0}$ is the part arising
from the contribution of (\ref{1}) related to the preparation in the
time interval $[T,{t_0}]$ where operators ${\hat
  a}^{\scriptscriptstyle (1)}_{ n}$, ${\hat a}^{\scriptscriptstyle
  (1){\scriptscriptstyle\dagger}}_{ n}$ and ${\hat
  a}^{\scriptscriptstyle (2)}_{ n}$, ${\hat a}^{\scriptscriptstyle
  (2){\scriptscriptstyle\dagger}}_{ n}$ appear together. An expansion
of ${\hat \varrho}_{t_0}$ with respect to ${\hat
  C}^{\scriptscriptstyle (12)}_{t_0}$ can be given, such that the
basic positivity property of ${\hat \varrho}_{t_0}$ is granted.
Starting from the representation:
\begin{displaymath}
e^{{\hat A}+{\hat B}}= 
\left({\hat 1}+        \int_0^1 du \,
        e^{u ({\hat A}+{\hat B}) }
        {\hat B}
        e^{-u{\hat A}}
\right)  
e^{{\hat A}}  
=
e^{{\hat A}}  
\left({\hat 1}+        \int_0^1 du \,
e^{-u{\hat A}}   {\hat B}     e^{u ({\hat A}+{\hat B}) }
\right)  
\end{displaymath}
one has the identity:
\begin{displaymath}
e^{{\hat A}+{\hat B}}=e^{{\hat A}+{\hat B}\over 2}e^{{\hat A}+{\hat
    B}\over 2}=   
\left({\hat 1}+        \int_0^{\frac{1}{2}} du \,
        e^{u ({\hat A}+{\hat B}) }
        {\hat B}
        e^{-u{\hat A}}
\right)  
e^{{\hat A}}  
\left({\hat 1}+        \int_0^{\frac{1}{2}} du \,
e^{-u{\hat A}}   {\hat B}     e^{u ({\hat A}+{\hat B}) }
\right)  
\end{displaymath}
which leads to the following expansion with respect to ${\hat B}$:
\begin{displaymath}
e^{{\hat A}+{\hat B}}=
\left({\hat 1}+        \int_0^{\frac{1}{2}} du \,
        e^{u {\hat A} }
        {\hat B}
        e^{-u{\hat A}}
+\ldots
\right)  
e^{{\hat A}}  
\left({\hat 1}+        \int_0^{\frac{1}{2}} du \,
e^{-u{\hat A}}   {\hat B}     e^{u {\hat A} }
+\ldots
\right).  
\end{displaymath}
The first term of this kind of expansion gives an insight into the
dynamics described by ${\hat \varrho}_{t_0}$, in fact setting
\begin{displaymath}
{\hat \varrho}^{\scriptscriptstyle (1)}_{t_0}={
\exp {\{- {\hat S}^{\scriptscriptstyle (1)}_{t_0} \}}
\over
\textrm{Tr}\left[
\exp {\{- {\hat S}^{\scriptscriptstyle (1)}_{t_0} \}}
\right]}
,
\qquad
{\hat \varrho}^{\scriptscriptstyle (2)}_{t_0}={
\exp {\{- {\hat S}^{\scriptscriptstyle (2)}_{t_0} \}}
\over
\textrm{Tr}\left[
\exp {\{- {\hat S}^{\scriptscriptstyle (2)}_{t_0} \}}
\right]}
\end{displaymath}
and
\begin{equation}
  \label{*}
  {\hat {\bf \textsf{C}}}^{\scriptscriptstyle (12)}_{t_0}=
  \int_0^{\frac{1}{2}} du \, 
        ({\hat \varrho}^{\scriptscriptstyle (1)}_{t_0}\otimes {\hat
          \varrho}^{\scriptscriptstyle (2)}_{t_0})^{u} 
        {\hat C}^{\scriptscriptstyle (12)}_{t_0}
        ({\hat \varrho}^{\scriptscriptstyle (1)}_{t_0}\otimes {\hat
          \varrho}^{\scriptscriptstyle (2)}_{t_0})^{-u}
\end{equation}
one has:
\begin{displaymath}
  {\hat \varrho}_{t_0} =
{
\left({\hat 1}+ 
        {\hat {\bf \textsf{C}}}^{\scriptscriptstyle (12)}_{t_0}
      \right)
{\hat \varrho}^{\scriptscriptstyle (1)}_{t_0}\otimes {\hat
  \varrho}^{\scriptscriptstyle (2)}_{t_0}  
\left({\hat 1}+         {\hat {\bf \textsf{C}}}^{\scriptscriptstyle
    (12){\scriptscriptstyle\dagger}}_{t_0}
      \right)
\over
\textrm{Tr}
\left[{
\left(
{\hat 1}+
        {\hat {\bf \textsf{C}}}^{\scriptscriptstyle (12)}_{t_0}
      \right)
{\hat \varrho}^{\scriptscriptstyle (1)}_{t_0}\otimes {\hat
  \varrho}^{\scriptscriptstyle (2)}_{t_0}  
\left(
{\hat 1}+ 
        {\hat {\bf \textsf{C}}}^{\scriptscriptstyle
          (12){\scriptscriptstyle\dagger}}_{t_0} 
\right)
}\right]}
.
\end{displaymath}
Let us consider
\begin{displaymath}
\hat{N}_1 =
  \int_{\omega_1} d^3\!{\mbox{\bf x}}\,
{\hat \psi}^{\scriptscriptstyle
  (1){\scriptscriptstyle\dagger}}({\mbox{\bf x}}){\hat
  \psi}^{\scriptscriptstyle 
  (1)}({\mbox{\bf x}}) 
=
\sum_n {\hat
  a}^{\scriptscriptstyle (1){\scriptscriptstyle\dagger}}_{ n}{\hat
  a}^{\scriptscriptstyle (1)}_{ n},
\ 
\hat{N}_2 =
  \int_{\omega_2} d^3\!{\mbox{\bf x}}\,
{\hat \psi}^{\scriptscriptstyle (2){\scriptscriptstyle
  \dagger}}({\mbox{\bf x}}){\hat \psi}^{\scriptscriptstyle
  (2)}({\mbox{\bf x}}) 
=
\sum_n {\hat
  a}^{\scriptscriptstyle (2){\scriptscriptstyle\dagger}}_{ n}{\hat
  a}^{\scriptscriptstyle (2)}_{ n}.
\end{displaymath}
The part of the operator ${\hat {\bf \textsf{C}}}^{\scriptscriptstyle
  (12)}_{t_0}$ 
that will have the 
most important role in our treatment has the form:
\begin{equation}
  \label{8}
  {\hat {\bf \textsf{C}}}^{\scriptscriptstyle (12)}_{t_0}=\sum_h {\hat
  a}^{\scriptscriptstyle (1){\scriptscriptstyle\dagger}}_{h}{\hat
  D}^{\scriptscriptstyle (2)}_{h} +\ldots 
\end{equation}
where ${\hat
  D}^{\scriptscriptstyle (2)}_{h}$ indicates an operator built with 
${\hat a}^{\scriptscriptstyle (2)}_{ n}$, ${\hat
  a}^{\scriptscriptstyle (2){\scriptscriptstyle\dagger}}_{ n}$ having
the role 
of a destruction operator in region 2, i.e.:
\begin{displaymath}
  \hat{N}_2{\hat
  D}^{\scriptscriptstyle (2)}_{h} ={\hat
  D}^{\scriptscriptstyle (2)}_{h} (\hat{N}_2 -1).
\end{displaymath}
In fact in expression (\ref{8}) only one creation operator ${\hat
  a}^{\scriptscriptstyle (1){\scriptscriptstyle\dagger}}_{h}$ related
to the modes of system 1 
appears, so that the remaining part related to system 2 must have the
role of a destruction operator for the system 2, since
\begin{displaymath}
  [{\hat {\bf \textsf{C}}}^{\scriptscriptstyle (12)}_{t_0},\hat{N}]=0,
\end{displaymath}
and all relevant variables in this non relativistic theory must
commute with $\hat{N}=\hat{N}_1+\hat{N}_2$.  Other terms would
be:
\begin{equation}
  \label{100}
\sum_{h_1, h_2} {\hat
  a}^{\scriptscriptstyle (1){\scriptscriptstyle\dagger}}_{h_1} {\hat
  a}^{\scriptscriptstyle (2){\scriptscriptstyle\dagger}}_{h_2}{\hat
  D}^{\scriptscriptstyle (2)}_{h_1, h_2}
,
\qquad
\hat{N}_2{\hat
  D}^{\scriptscriptstyle (2)}_{h_1, h_2}={\hat
  D}^{\scriptscriptstyle (2)}_{h_1, h_2}   (\hat{N}_2 -2),
\end{equation}
and similar ones. These kind of contributions would simply be
associated in our treatment with a ``two-particle'' system, or more
generally a compound microsystem. Also contributions in which the role
of the indexes 1 and 2 is exchanged generally arise (however not
necessarily since ${\hat {\bf \textsf{C}}}^{\scriptscriptstyle
  (12)}_{t_0}$, differently from 
${\hat C}^{\scriptscriptstyle (12)}_{t_0}$, does not need to be
self-adjoint), such terms 
would have a similar physical interpretation simply by interchanging
the role of system 1 and system 2. However also other terms generally
appear together with contribution (\ref{8}), e.g.
\begin{displaymath}
\sum_{h_1, h_2,k} {\hat
  a}^{\scriptscriptstyle (1){\scriptscriptstyle\dagger}}_{h_1} {\hat
  a}^{\scriptscriptstyle (1){\scriptscriptstyle\dagger}}_{h_2}{\hat
  a}^{\scriptscriptstyle (1)}_{k}{\hat
  D}^{\scriptscriptstyle (2)}_{h_1, h_2,k}
+
\sum_{h_1, h_2,h_3,k_1,k_2} {\hat
  a}^{\scriptscriptstyle (1){\scriptscriptstyle\dagger}}_{h_1} {\hat
  a}^{\scriptscriptstyle (1){\scriptscriptstyle\dagger}}_{h_2}{\hat
  a}^{\scriptscriptstyle (1){\scriptscriptstyle\dagger}}_{h_3}{\hat
  a}^{\scriptscriptstyle (1)}_{k_1}{\hat
  a}^{\scriptscriptstyle (1)}_{k_2}{\hat
  D}^{\scriptscriptstyle (2)}_{h_1, h_2, h_3,k_1,k_2}
+
\ldots.
\end{displaymath}
In fact already due to the
structure (\ref{*}) of ${\hat {\bf \textsf{C}}}^{\scriptscriptstyle
  (12)}_{t_0}$ and the correlations among the modes
$u_n^{\scriptscriptstyle (1)}$ in ${\hat \varrho}^{\scriptscriptstyle
  (1)}_{t_0}$ these terms would anyway arise in ${\hat {\bf
    \textsf{C}}}^{\scriptscriptstyle (12)}_{t_0}$. However we must
expect that 
for meaningful preparation procedures operators ${\hat {\bf
    \textsf{C}}}^{\scriptscriptstyle (12)}_{t_0}$ generally arise
whose 
contribution to the dynamics of relevant variables decays for time
$t>t_0+\tau$, $\tau$ being a microscopic correlation time: otherwise
any treatment of non-equilibrium isolated systems would be impossible.
Indeed the very scope of our article is to indicate a very particular
situation in which this general assumption fails. So it seems
reasonable to assume that the contribution of these terms to the
dynamics of relevant variables can be neglected for $t\geq t_0+\tau$.
Clearly this point requires further investigation. In conclusion we
shall now investigate the dynamics of ${\hat \varrho}_t$ for $t\geq
t_0+\tau$, taking simply
\begin{displaymath}
  {\hat {\bf \textsf{C}}}^{\scriptscriptstyle (12)}_{t_0}=\sum_h {\hat
  a}^{\scriptscriptstyle (1){\scriptscriptstyle\dagger}}_{h}{\hat
  D}^{\scriptscriptstyle (2)}_{h}
\end{displaymath}
so that
\begin{displaymath}
{\hat \varrho}_t
=
{
{\cal U}^{\scriptscriptstyle (1)}_{t-t_0}\otimes{\cal
  U}^{\scriptscriptstyle (2)}_{t-t_0} 
[
{\hat 1}+ 
\sum_h {\hat
  a}^{\scriptscriptstyle (1){\scriptscriptstyle\dagger}}_{h}{\hat
  D}^{\scriptscriptstyle (2)}_{h}
]
{\hat \varrho}^{\scriptscriptstyle (1)}_{t_0}\otimes {\hat
          \varrho}^{\scriptscriptstyle (2)}_{t_0}
[
{\hat 1}+ 
\sum_k {\hat
  a}^{\scriptscriptstyle (1)}_{k}{\hat
  D}^{\scriptscriptstyle (2){\scriptscriptstyle\dagger}}_{k}
]
{\cal U}^{\scriptscriptstyle (1){\scriptscriptstyle\dagger}}_{t-t_0}
\otimes{\cal 
  U}^{\scriptscriptstyle (2){\scriptscriptstyle\dagger}}_{t-t_0} 
\over
\textrm{Tr}\left[
[
{\hat 1}+ 
\sum_h {\hat
  a}^{\scriptscriptstyle (1){\scriptscriptstyle\dagger}}_{h}{\hat
  D}^{\scriptscriptstyle (2)}_{h}
]
{\hat \varrho}^{\scriptscriptstyle (1)}_{t_0}\otimes {\hat
          \varrho}^{\scriptscriptstyle (2)}_{t_0}
[
{\hat 1}+ 
\sum_k {\hat
  a}^{\scriptscriptstyle (1)}_{k}{\hat
  D}^{\scriptscriptstyle (2){\scriptscriptstyle\dagger}}_{k}
]
\right]}
\end{displaymath}
for $t-t_0 >\tau$.
Let us stress that for all observables $\hat{A}={\hat
  A}^{\scriptscriptstyle (1)}\otimes {\hat A}^{\scriptscriptstyle
  (2)}$ such that
\begin{equation}
\label{11}
[{\hat A}^{\scriptscriptstyle (1)},\hat{N}_1]=0,\quad [{\hat
  A}^{\scriptscriptstyle (2)},\hat{N}_2]=0, 
\end{equation}
and these are the only meaningful observables for systems 1 and 2,
${\hat \varrho}_{t}$ is equivalent to the mixture
\begin{eqnarray}
  \label{l}
  {\hat
  \varrho}_{t}&=&
\lambda\,
{\cal U}^{\scriptscriptstyle (1)}_{t-t_0}\otimes{\cal
  U}^{\scriptscriptstyle (2)}_{t-t_0} \,
{\hat \varrho}^{\scriptscriptstyle (1)}_{t_0}\otimes {\hat
          \varrho}^{\scriptscriptstyle (2)}_{t_0}\,
{\cal U}^{\scriptscriptstyle
  (1){\scriptscriptstyle\dagger}}_{t-t_0}\otimes{\cal 
  U}^{\scriptscriptstyle (2){\scriptscriptstyle\dagger}}_{t-t_0}  
\\
\nonumber
&&{}+
(1-\lambda)
{
{\cal U}^{\scriptscriptstyle (1)}_{t-t_0}\otimes{\cal
  U}^{\scriptscriptstyle (2)}_{t-t_0} 
\,\sum_h {\hat
  a}^{\scriptscriptstyle (1){\scriptscriptstyle\dagger}}_{h}{\hat
  D}^{\scriptscriptstyle (2)}_{h}
\,{\hat \varrho}^{\scriptscriptstyle (1)}_{t_0}\otimes {\hat
          \varrho}^{\scriptscriptstyle (2)}_{t_0}
\,\sum_k {\hat
  a}^{\scriptscriptstyle (1)}_{k}{\hat
  D}^{\scriptscriptstyle (2){\scriptscriptstyle\dagger}}_{k}
{\cal U}^{\scriptscriptstyle
  (1){\scriptscriptstyle\dagger}}_{t-t_0}\otimes{\cal 
  U}^{\scriptscriptstyle (2){\scriptscriptstyle\dagger}}_{t-t_0}
\over
\textrm{Tr}\left[
\,\sum_h {\hat
  a}^{\scriptscriptstyle (1){\scriptscriptstyle\dagger}}_{h}{\hat
  D}^{\scriptscriptstyle (2)}_{h}
\,{\hat \varrho}^{\scriptscriptstyle (1)}_{t_0}\otimes {\hat
          \varrho}^{\scriptscriptstyle (2)}_{t_0}
\,\sum_k {\hat
  a}^{\scriptscriptstyle (1)}_{k}{\hat
  D}^{\scriptscriptstyle (2){\scriptscriptstyle\dagger}}_{k}
\right]}
\end{eqnarray}
where 
\begin{displaymath}
  \lambda^{-1}=\textrm{Tr}\left[
\sum_h {\hat
  a}^{\scriptscriptstyle (1){\scriptscriptstyle\dagger}}_{h}{\hat
  D}^{\scriptscriptstyle (2)}_{h}
\,{\hat \varrho}^{\scriptscriptstyle (1)}_{t_0}\otimes {\hat
          \varrho}^{\scriptscriptstyle (2)}_{t_0}
\,\sum_k {\hat
  a}^{\scriptscriptstyle (1)}_{k}{\hat
  D}^{\scriptscriptstyle (2){\scriptscriptstyle\dagger}}_{k}
\right]
.
\end{displaymath}
In fact one has
\begin{displaymath}
  \textrm{Tr}\left[
{\hat A}^{\scriptscriptstyle (1)}\otimes {\hat A}^{\scriptscriptstyle
  (2)} \,\sum_h {\hat 
  a}^{\scriptscriptstyle (1){\scriptscriptstyle\dagger}}_{h}{\hat
  D}^{\scriptscriptstyle (2)}_{h}\,{\hat \varrho}^{\scriptscriptstyle
  (1)}_{t_0}\otimes {\hat 
          \varrho}^{\scriptscriptstyle (2)}_{t_0}\right]=0,
\end{displaymath}
as can be seen indicating with $|N_1,N_2,S\rangle$ a basis of
eigenstates of $\hat{N}_1$, $\hat{N}_2$, so that:
\begin{eqnarray*}
&&   \textrm{Tr}\left[
{\hat A}^{\scriptscriptstyle (1)}\otimes {\hat A}^{\scriptscriptstyle
  (2)}\, \sum_h {\hat 
  a}^{\scriptscriptstyle (1){\scriptscriptstyle\dagger}}_{h}{\hat
  D}^{\scriptscriptstyle (2)}_{h}\,{\hat \varrho}^{\scriptscriptstyle
  (1)}_{t_0}\otimes {\hat 
          \varrho}^{\scriptscriptstyle (2)}_{t_0}\right]=
\\
&&
\sum_{N_1,N_2,S}
\langle N_1,N_2,S |{\hat A}^{\scriptscriptstyle (1)}\otimes {\hat
  A}^{\scriptscriptstyle (2)} \,\sum_h {\hat 
  a}^{\scriptscriptstyle (1){\scriptscriptstyle\dagger}}_{h}{\hat
  D}^{\scriptscriptstyle (2)}_{h}\,{\hat \varrho}^{\scriptscriptstyle
  (1)}_{t_0}\otimes {\hat 
          \varrho}^{\scriptscriptstyle (2)}_{t_0} |N_1,N_2,S\rangle
=0 
\end{eqnarray*}
since due to (\ref{11}) the state
\begin{displaymath}
{\hat A}^{\scriptscriptstyle (1)}\otimes {\hat
  A}^{\scriptscriptstyle (2)} \,\sum_h {\hat 
  a}^{\scriptscriptstyle (1){\scriptscriptstyle\dagger}}_{h}{\hat
  D}^{\scriptscriptstyle (2)}_{h}\,{\hat \varrho}^{\scriptscriptstyle
  (1)}_{t_0}\otimes {\hat 
          \varrho}^{\scriptscriptstyle (2)}_{t_0} |N_1,N_2,S\rangle
\end{displaymath}
belongs to the eigenstate of $\hat{N}_1$, $\hat{N}_2$ corresponding to
eigenvalue $N_1+1$, $N_2 -1$.  Let us comment on the physical meaning
of (\ref{l}). Decomposing the expectations of the relevant variables
according to this mixture one can associate to the two components of
the mixture different state parameters $\zeta (t)$, different
corresponding Gibbs states and different entropies. For the dynamics
of the first component of the mixture, constructing in the usual way
the non-equilibrium statistical operator, it is possible to assume
that a ``simple dynamics'' arises, i.e., the memory term in the
non-equilibrium statistical operator decays within a correlation time:
then the usual methods of non-equilibrium statistical mechanics apply.
If we had considered terms like (\ref{100}) and also terms in which
the role of indexes 1 and 2 is reversed, we would still obtain ${\hat
  \varrho}_{t}$ equivalent to a mixture of statistical operators
corresponding to these different situations, with an obvious physical
meaning.  For the dynamics of the second component the typical
structure of the non-equilibrium statistical operator is no longer
feasible. We shall see that the dynamics of a ``microsystem'' becomes
an important ingredient to describe what happens.

\section{Initial correlation and the appearance of a microsystem}
The dynamics of the second component of the mixture (\ref{l}) is given by
\begin{eqnarray}
  \label{2.1}
  {\hat
  \varrho}_{t}
  &=&
{
{\cal U}^{\scriptscriptstyle (1)}_{t-t_0}\otimes{\cal
  U}^{\scriptscriptstyle (2)}_{t-t_0} 
\,\sum_h {\hat
  a}^{\scriptscriptstyle (1){\scriptscriptstyle\dagger}}_{h}{\hat
  D}^{\scriptscriptstyle (2)}_{h}
{\hat \varrho}^{\scriptscriptstyle (1)}_{t_0}\otimes {\hat
          \varrho}^{\scriptscriptstyle (2)}_{t_0}
\,\sum_k {\hat
  a}^{\scriptscriptstyle (1)}_{k}{\hat
  D}^{\scriptscriptstyle (2){\scriptscriptstyle\dagger}}_{k}
{\cal U}^{\scriptscriptstyle
  (1){\scriptscriptstyle\dagger}}_{t-t_0}\otimes{\cal 
  U}^{\scriptscriptstyle (2){\scriptscriptstyle\dagger}}_{t-t_0}
\over
\textrm{Tr}\left[
\,\sum_h {\hat
  a}^{\scriptscriptstyle (1){\scriptscriptstyle\dagger}}_{h}{\hat
  D}^{\scriptscriptstyle (2)}_{h}
{\hat \varrho}^{\scriptscriptstyle (1)}_{t_0}\otimes {\hat
          \varrho}^{\scriptscriptstyle (2)}_{t_0}
\,\sum_k {\hat
  a}^{\scriptscriptstyle (1)}_{k}{\hat
  D}^{\scriptscriptstyle (2){\scriptscriptstyle\dagger}}_{k}
\right]}  
\\
\nonumber
&=&
{
\sum_{h,k}
{\cal U}^{\scriptscriptstyle (1)}_{t-t_0}
{\hat
  a}^{\scriptscriptstyle (1){\scriptscriptstyle\dagger}}_{h}
{\cal U}^{\scriptscriptstyle (1){\scriptscriptstyle\dagger}}_{t-t_0}
{\hat \varrho}^{\scriptscriptstyle (1)}_{t}
{\cal U}^{\scriptscriptstyle (1)}_{t-t_0}
{\hat
  a}^{\scriptscriptstyle (1)}_{k}
{\cal U}^{\scriptscriptstyle (1){\scriptscriptstyle\dagger}}_{t-t_0}
\otimes{\cal U}^{\scriptscriptstyle (2)}_{t-t_0}
{\hat
  D}^{\scriptscriptstyle (2)}_{h}
{\hat \varrho}^{\scriptscriptstyle (2)}_{t_0}
{\hat
  D}^{\scriptscriptstyle (2){\scriptscriptstyle\dagger}}_{k}
{\cal
  U}^{\scriptscriptstyle (2){\scriptscriptstyle\dagger}}_{t-t_0}
\over
\textrm{Tr}\left[
\,\sum_h {\hat
  a}^{\scriptscriptstyle (1){\scriptscriptstyle\dagger}}_{h}{\hat
  D}^{\scriptscriptstyle (2)}_{h}
{\hat \varrho}^{\scriptscriptstyle (1)}_{t_0}\otimes {\hat
          \varrho}^{\scriptscriptstyle (2)}_{t_0}
\,\sum_k {\hat
  a}^{\scriptscriptstyle (1)}_{k}{\hat
  D}^{\scriptscriptstyle (2){\scriptscriptstyle\dagger}}_{k}
\right]}
.  
\end{eqnarray}
According to the fact that we have chosen the term (\ref{8}) to
characterize the structure of ${\hat {\bf \textsf{C}}}^{\scriptscriptstyle
  (12)}_{t_0}$, we are now mainly interested in the dynamics of
observables ${\hat A}^{\scriptscriptstyle (1)}\otimes{\hat
  1}^{\scriptscriptstyle (2)}$, then one has:
\begin{eqnarray}
  \label{2.2}
&&
  \textrm{Tr}\left[ {\hat
  A}^{\scriptscriptstyle (1)}\otimes{\hat 1}^{\scriptscriptstyle
  (2)}{\hat \varrho}_t \right]= 
\\
\nonumber
&&
\hphantom{sposta}
{
\sum_{h,k}
\textrm{Tr}_{{\cal H}^{\scriptscriptstyle (1)}}\left[
{\hat A}^{\scriptscriptstyle (1)}
{\cal U}^{\scriptscriptstyle (1)}_{t-t_0}
{\hat
  a}^{\scriptscriptstyle (1){\scriptscriptstyle\dagger}}_{h}
{\cal U}^{\scriptscriptstyle (1){\scriptscriptstyle\dagger}}_{t-t_0}
{\hat \varrho}^{\scriptscriptstyle (1)}_{t}
{\cal U}^{\scriptscriptstyle (1)}_{t-t_0}
{\hat
  a}^{\scriptscriptstyle (1)}_{k}
{\cal U}^{\scriptscriptstyle
  (1){\scriptscriptstyle\dagger}}_{t-t_0}
\right]
\textrm{Tr}_{{\cal H}^{\scriptscriptstyle (2)}}\left[
{\hat
  D}^{\scriptscriptstyle (2)}_{h}
{\hat \varrho}^{\scriptscriptstyle (2)}_{t_0}
{\hat
  D}^{\scriptscriptstyle (2){\scriptscriptstyle\dagger}}_{k}
\right]\over
\sum_{h,k}
\textrm{Tr}_{{\cal H}^{\scriptscriptstyle (1)}}\left[
{\hat
  a}^{\scriptscriptstyle (1){\scriptscriptstyle\dagger}}_{h}
{\hat \varrho}^{\scriptscriptstyle (1)}_{t_0}
{\hat
  a}^{\scriptscriptstyle (1)}_{k}\right]
\textrm{Tr}_{{\cal H}^{\scriptscriptstyle (2)}}\left[
{\hat
  D}^{\scriptscriptstyle (2)}_{h}
{\hat \varrho}^{\scriptscriptstyle (2)}_{t_0}
{\hat
  D}^{\scriptscriptstyle (2){\scriptscriptstyle\dagger}}_{k}
\right]
}
.
\end{eqnarray}
Let us introduce the following notations:
\begin{eqnarray*}
  w_{hk} (t_0)&=&
{
\textrm{Tr}_{{\cal H}^{\scriptscriptstyle (2)}}\left[
{\hat
  D}^{\scriptscriptstyle (2)}_{h}
{\hat \varrho}^{\scriptscriptstyle (2)}_{t_0}
{\hat
  D}^{\scriptscriptstyle (2){\scriptscriptstyle\dagger}}_{k}
\right]
\over
\sum_h
\textrm{Tr}_{{\cal H}^{\scriptscriptstyle (2)}}\left[
{\hat
  D}^{\scriptscriptstyle (2){\scriptscriptstyle\dagger}}_{h}
{\hat
  D}^{\scriptscriptstyle (2)}_{h}
{\hat \varrho}^{\scriptscriptstyle (2)}_{t_0}
\right]}
\\
\nonumber
A_{kh} (t)&=&
\textrm{Tr}_{{\cal H}^{\scriptscriptstyle (1)}}\left[
{\hat A}^{\scriptscriptstyle (1)}
{\cal U}^{\scriptscriptstyle (1)}_{t-t_0}
{\hat
  a}^{\scriptscriptstyle (1){\scriptscriptstyle\dagger}}_{h}
{\cal U}^{\scriptscriptstyle (1){\scriptscriptstyle\dagger}}_{t-t_0}
{\hat \varrho}^{\scriptscriptstyle (1)}_{t}
{\cal U}^{\scriptscriptstyle (1)}_{t-t_0}
{\hat
  a}^{\scriptscriptstyle (1)}_{k}
{\cal U}^{\scriptscriptstyle (1){\scriptscriptstyle\dagger}}_{t-t_0}
\right],
\end{eqnarray*}
then (\ref{2.2}) can be written as
\begin{equation}
  \label{2.3}
  \textrm{Tr}\left[ {\hat
  A}^{\scriptscriptstyle (1)}\otimes{\hat 1}^{\scriptscriptstyle
  (2)}{\hat \varrho}_t \right]= 
\sum_{h,k} A_{kh} (t)  w_{hk} (t_0)
{
\sum_h
\textrm{Tr}_{{\cal H}^{\scriptscriptstyle (2)}}\left[
{\hat
  D}^{\scriptscriptstyle (2){\scriptscriptstyle\dagger}}_{h}
{\hat
  D}^{\scriptscriptstyle (2)}_{h}
{\hat \varrho}^{\scriptscriptstyle (2)}_{t_0}
\right]
\over
\sum_{h,k}
\textrm{Tr}_{{\cal H}^{\scriptscriptstyle (1)}}\left[
{\hat
  a}^{\scriptscriptstyle (1){\scriptscriptstyle\dagger}}_{h}
{\hat \varrho}^{\scriptscriptstyle (1)}_{t_0}
{\hat
  a}^{\scriptscriptstyle (1)}_{k}\right]
\textrm{Tr}_{{\cal H}^{\scriptscriptstyle (2)}}\left[
{\hat
  D}^{\scriptscriptstyle (2)}_{h}
{\hat \varrho}^{\scriptscriptstyle (2)}_{t_0}
{\hat
  D}^{\scriptscriptstyle (2){\scriptscriptstyle\dagger}}_{k}
\right]}
\end{equation}
and hints towards the typical structure of one-particle
quantum mechanics. To $w_{hk} (t_0)$ the role can be given of the
statistical operator at time $t_0$ for a one-particle system, when one
takes as a basis to represent it the normal modes of system 1 in $L^2
(\omega_1)$; $A_{kh} (t)$ can be seen as the representative of an
observable, embodying time dependence. Obviously we will have to check
this idea looking at the dynamics of the system: in this way the
correlation between system 1 and system 2 should be explained in terms
of a particle produced by system 2 and detected by system 1. However
all this holds if the additional correlation described by the
numerical factor
\begin{displaymath}
  \sigma={
\sum_h
\textrm{Tr}_{{\cal H}^{\scriptscriptstyle (2)}}\left[
{\hat
  D}^{\scriptscriptstyle (2){\scriptscriptstyle\dagger}}_{h}
{\hat
  D}^{\scriptscriptstyle (2)}_{h}
{\hat \varrho}^{\scriptscriptstyle (2)}_{t_0}
\right]\over
\sum_{h,k}
\textrm{Tr}_{{\cal H}^{\scriptscriptstyle (1)}}\left[
{\hat
  a}^{\scriptscriptstyle (1){\scriptscriptstyle\dagger}}_{h}
{\hat \varrho}^{\scriptscriptstyle (1)}_{t_0}
{\hat
  a}^{\scriptscriptstyle (1)}_{k}\right]
\textrm{Tr}_{{\cal H}^{\scriptscriptstyle (2)}}\left[
{\hat
  D}^{\scriptscriptstyle (2)}_{h}
{\hat \varrho}^{\scriptscriptstyle (2)}_{t_0}
{\hat
  D}^{\scriptscriptstyle (2){\scriptscriptstyle\dagger}}_{k}
\right]}
\end{displaymath}
can be neglected. We shall see that by rather natural assumptions this
factor is practically one. In fact let us diagonalize the matrix $w_{hk}
(t_0)$
\begin{displaymath}
  w_{hk} (t_0)=\sum_\alpha \lambda^\alpha (t_0) \langle
  h|\psi^\alpha_{0}\rangle\langle\psi^\alpha_{0} | k\rangle,
\quad
\lambda^\alpha (t_0) > 0,
\quad 
\sum_\alpha \lambda^\alpha (t_0) =1;
\end{displaymath}
then:
\begin{displaymath}
  \frac{1}{\sigma}=
\sum_{h,k,\alpha}
 \textrm{Tr}_{{\cal H}^{\scriptscriptstyle (1)}}\left[
{\hat
  a}^{\scriptscriptstyle (1){\scriptscriptstyle\dagger}}_{h}
{\hat \varrho}^{\scriptscriptstyle (1)}_{t_0}
{\hat
  a}^{\scriptscriptstyle (1)}_{k}\right]
 \langle
  h|\psi^\alpha_{0}\rangle\langle\psi^\alpha_{0} | k\rangle
\lambda^\alpha (t_0)
=
\sum_{\alpha}
 \textrm{Tr}_{{\cal H}^{\scriptscriptstyle (1)}}\left[
{\hat
  a}^{\scriptscriptstyle (1){\scriptscriptstyle\dagger}}_{\psi^\alpha_0}
{\hat \varrho}^{\scriptscriptstyle (1)}_{0}
{\hat a}^{\scriptscriptstyle (1)}_{\psi^\alpha_{0}}\right]
\lambda^\alpha (t_0)
\end{displaymath}
with
\begin{equation}
  \label{-}
  {\hat a}^{\scriptscriptstyle (1)}_{\psi^\alpha_{0}}
=
\sum_{k}{\hat
  a}^{\scriptscriptstyle (1)}_{k}
\langle\psi^\alpha_{0} | k\rangle,
\end{equation}
where ${\hat a}^{\scriptscriptstyle (1)}_{\psi^\alpha_{0}}$ can be
considered 
as the annihilation operator of a particle in the state $\sum_{h}
u_h(\textbf{x})\langle h|\psi^\alpha_{0}\rangle$. Let us assume for
the moment that the statistical operator $w(t_0)$ is a pure state
$w(t_0)=|\psi_{0}\rangle\langle\psi_{0} |$, then:
\begin{displaymath}
      \frac{1}{\sigma}=
 \textrm{Tr}_{{\cal H}^{\scriptscriptstyle (1)}}\left[
{\hat a}^{\scriptscriptstyle (1){\scriptscriptstyle\dagger}}_{\psi_{0}}
{\hat \varrho}^{\scriptscriptstyle (1)}_{t_0}
{\hat a}^{\scriptscriptstyle (1)}_{\psi_{0}}\right]
=
1\pm 
 \textrm{Tr}_{{\cal H}^{\scriptscriptstyle (1)}}\left[
{\hat a}^{\scriptscriptstyle (1){\scriptscriptstyle\dagger}}_{\psi_{0}}
{\hat a}^{\scriptscriptstyle (1)}_{\psi_{0}}
{\hat \varrho}^{\scriptscriptstyle (1)}_{t_0}\right]
\end{displaymath}
with ${\hat a}^{\scriptscriptstyle (1)}_{\psi_{0}}=\sum_{k}{\hat
  a}^{\scriptscriptstyle (1)}_{k} \langle\psi_{0} | k\rangle$.
Taking into account that system 1 should have the role of a detection
device for the microsystem it is rather natural to assume that this
one-particle state $|\psi_{0}\rangle$ is very scarcely occupied,
i.e.:
\begin{equation}
  \label{2.5}
   \textrm{Tr}_{{\cal H}^{\scriptscriptstyle (1)}}\left[
{\hat a}^{\scriptscriptstyle (1){\scriptscriptstyle\dagger}}_{\psi_{0}}
{\hat a}^{\scriptscriptstyle (1)}_{\psi_{0}}
{\hat \varrho}^{\scriptscriptstyle (1)}_{t_0}\right] \ll 1,
\end{equation}
and as a consequence $\sigma\approx 1$. More generally we shall assume
that $w(t_0)$ is a mixture of one-particle states $\{
|\psi^\alpha_{0}\rangle 
\ \alpha=1,2,\ldots,n \}$ that are scarcely occupied. Let us stress that it
is just the skill of the experimentalist to build sources that prepare
possibly pure states. In this situation (\ref{2.3}) becomes:
\begin{displaymath}
  \textrm{Tr}\left[ {\hat
  A}^{\scriptscriptstyle (1)}\otimes{\hat 1}^{\scriptscriptstyle
  (2)}{\hat \varrho}_t \right]= 
\sum_{h,k} A_{kh} (t)  w_{hk} (t_0)
=
\textrm{Tr}_{L^2
(\omega_1)}\left[
A (t)w (t_0)\right],  
\end{displaymath}
where in the last term the trace refers to the one-particle Hilbert space
in which the operators $A (t)$, $w (t_0)$ are represented. Let us
observe that the correspondence from ${\hat A}^{\scriptscriptstyle
  (1)}$ to $A (t)$ has the following properties: ${\hat
  1}^{\scriptscriptstyle (1)}$ is transferred to the identity on $L^2
(\omega_1)$, positivity is preserved and therefore a projection valued
or p.o.v. measure ${\hat E}^{\scriptscriptstyle (1)}$ is transferred
to a p.o.v. measure on $L^2 (\omega_1)$. Let us consider a little
further the dynamics of the macrosystem 1 referring to the
subcollection that has to be demixed at time $t_0+\tau$ and described
by the statistical operator (\ref{2.1}). Let us restrict for
simplicity to the case in which the source prepares a pure state
$\psi_{0}$. Then one has from (\ref{2.3}) the following expression
for expectations of the observables of system 1 at a time $t>t_0+\tau$:
\begin{displaymath}
  \langle {\hat
  A}^{\scriptscriptstyle (1)} \rangle_t=
\textrm{Tr}_{{\cal H}^{\scriptscriptstyle (1)}}\left[
{\hat A}^{\scriptscriptstyle (1)}
{\cal U}^{\scriptscriptstyle (1)}_{t-t_0}
{\hat a}_{\psi_{0}}^{\scriptscriptstyle\dagger}
{\cal U}^{\scriptscriptstyle (1){\scriptscriptstyle\dagger}}_{t-t_0}
{\hat \varrho}^{\scriptscriptstyle (1)}_{t}
{\cal U}^{\scriptscriptstyle (1)}_{t-t_0}
{\hat  a}_{\psi_{0}}
{\cal U}^{\scriptscriptstyle
  (1){\scriptscriptstyle\dagger}}_{t-t_0}\right] 
=
\textrm{Tr}_{{\cal H}^{\scriptscriptstyle (1)}}\left[
{\hat A}^{\scriptscriptstyle (1)} (t)
{\hat \varrho}^{\scriptscriptstyle (1)}_{t}\right]
\end{displaymath}
where
\begin{displaymath}
{\hat A}^{\scriptscriptstyle (1)} (t)={\hat  a}_{\psi_0} (t){\hat
  A}^{\scriptscriptstyle (1)}{\hat 
  a}_{\psi_0}^{\scriptscriptstyle\dagger} (t)
\end{displaymath}
with
\begin{displaymath}
{\hat  a}_{\psi_0} (t)= {\cal U}^{\scriptscriptstyle (1)}_{t-t_0}
{\hat a}_{\psi_0}^{\scriptscriptstyle\dagger}
{\cal U}^{\scriptscriptstyle (1){\scriptscriptstyle\dagger}}_{t-t_0}.
\end{displaymath}
Then one is led to a redefinition of the relevant observables for the
macrosystem influenced by the microsystem
\begin{displaymath}
{\hat A}^{\scriptscriptstyle (1)}_j(\textbf{x}) \rightarrow 
{\hat {\bf \textsf{A}}}^{\scriptscriptstyle (1)}_j(\textbf{x})
\end{displaymath}
\begin{equation}
  \label{****}
{\hat {\bf \textsf{A}}}^{\scriptscriptstyle (1)}_j(\textbf{x})
 ={\hat  a}_{\psi_0} (t){\hat A}^{\scriptscriptstyle
   (1)}_j(\textbf{x}){\hat  
  a}_{\psi_0}^{\scriptscriptstyle\dagger} (t)=
{\hat A}^{\scriptscriptstyle (1)}_j(\textbf{x})+[{\hat  a}_{\psi_0}
(t),{\hat A}^{\scriptscriptstyle (1)}_j(\textbf{x})]{\hat 
  a}_{\psi_0}^{\scriptscriptstyle\dagger} (t)\pm {\hat
  A}^{\scriptscriptstyle (1)}_j(\textbf{x}){\hat 
  a}_{\psi_0}^{\scriptscriptstyle\dagger}(t){\hat  a}_{\psi_0} (t)
\end{equation}
and one considers a new reference Gibbs state related to ${\hat {\bf
    \textsf{A}}}_j(\textbf{x})$ and the corresponding state
parameters. Then one can study the dynamics of ${\hat
  \varrho}^{\scriptscriptstyle (1)}_{t}$ writing it as a new
equilibrium statistical operator for $t>t_0+\tau$ and a decisive
simplification would occur if its dynamics should be ``simple''. We
can fairly well expect that this is the case. The time dependence of
${\hat a}_{\psi_0} (t)$ is linked to the dynamics of the microsystem
as we shall discuss in the next sections: it will even give an insight
into the physics of the microsystem. Here we want to stress how the
concept of microsystem was helpful to face the problem of a non simple
dynamics of a macrosystem that we have here demixed into two possibly
simple dynamics. Let us observe that the time dependence of the
expectations of observables of system 1 is twofold: it arises once
from ${\hat a}_{\psi_0} (t)$ and is related to the physics of the
microsystem, it arises further from ${\hat
  \varrho}^{\scriptscriptstyle (1)}_{t}$ and is related to the general
dynamics of the macrosystem; one can easily characterize variables
that are scarcely effected by the microsystem. Let us consider ${\hat
  A}^{\scriptscriptstyle (1)}_j(\textbf{x})$ such that
\begin{displaymath}
  [{\hat  a}_{\psi_0} (t),{\hat A}^{\scriptscriptstyle
    (1)}_j(\textbf{x})]=0, 
\end{displaymath}
furthermore one can expect that
\begin{displaymath}
\textrm{Tr}\left[
{\hat A}^{\scriptscriptstyle (1)}_j(\textbf{x}){\hat
  a}_{\psi_0}^{\scriptscriptstyle\dagger}(t){\hat
  a}_{\psi_0} (t){\hat \varrho}^{\scriptscriptstyle (1)}_{t}\right]\ll
\textrm{Tr}\left[  
{\hat A}^{\scriptscriptstyle (1)}_j(\textbf{x}){\hat
  \varrho}^{\scriptscriptstyle 
  (1)}_{t}\right] 
\end{displaymath}
in fact it was assumed by (\ref{2.5}) that the state $\psi_0$ was
depleted for ${\hat \varrho}^{\scriptscriptstyle (1)}_{t_0}$, so that
this inequality should hold at time $t=t_0$ and one can expect that it
also holds during the time evolution of the microsystem in absence of
too strong interactions with the macrosystem. Then for these ${\hat
  A}_j(\textbf{x})$ one has by (\ref{****})
\begin{displaymath}
\langle{\hat {\bf \textsf{A}}}^{\scriptscriptstyle (1)}_j(\textbf{x})
\rangle_t
=
\textrm{Tr}_{{\cal H}^{\scriptscriptstyle (1)}}\left[
{\hat A}^{\scriptscriptstyle (1)}_j(\textbf{x})
{\hat \varrho}^{\scriptscriptstyle (1)}_{t}\right]
\end{displaymath}
and no dynamical consequence due to the microsystem arises. On the
contrary if by the choice of an observable ${\hat
  A}^{\scriptscriptstyle (1)}_j(\textbf{x})$ 
one meets the situation $[{\hat a}_{\psi_0} (t),{\hat
  A}_j(\textbf{x})]\not= 0$, one expects that by a suitable ${\hat
  \varrho}^{\scriptscriptstyle (1)}_{\bar t}$ a significant
macroscopic signal can arise: in this way the dynamics of the
microsystem in the time interval $[t_0,t]$ can be tested.

\section{Dynamics of the microsystem}
Let us first make the simple assumption that the interaction of
a particle created by ${\hat a}_{\psi_0}^{\scriptscriptstyle\dagger}$
with the different modes of the macrosystem can be neglected, then one
has immediately by (\ref{-})
\begin{displaymath}
{\hat
  a}_{\psi_0}^{\scriptscriptstyle\dagger}(t)=
{\cal U}_{t-t_0} {\hat
  a}_{\psi_0}^{\scriptscriptstyle\dagger}
{\cal U}_{t-t_0}^{\scriptscriptstyle\dagger} =
{\cal U}_{t-t_0} 
\sum_{h}{\hat
  a}_{h}^{\scriptscriptstyle\dagger}
\langle h | {\psi_0}\rangle
{\cal U}_{t-t_0}^{\scriptscriptstyle\dagger}
=
\sum_{h}
e^{-\frac{i}{\hbar}W_h (t-t_0)}
{\hat
  a}_{h}^{\scriptscriptstyle\dagger}
\langle h | {\psi_0}\rangle
=
{\hat
  a}_{\psi_{t}}^{\scriptscriptstyle\dagger}
\end{displaymath}
where $ {\psi_{t}}(\textbf{x})=\sum_{h} u_h(\textbf{x})
e^{-\frac{i}{\hbar}W_h (t-t_0) }\langle h|\psi_{0}\rangle $ is a
one-particle 
wave function, determined by the coefficients $\langle
h|\psi_{0}\rangle$ which are given by (\ref{-}) and were obtained
studying the source part related to system 2. ${\psi_{t}}(\textbf{x})$
satisfies the Schr\"odinger equation which rules the normal modes of
the Schr\"odinger field: then one-particle quantum mechanics is
extracted from the formalism. Of course this is only an approximation:
the microsystem interacts with the normal modes of the macrosystem and
the general problem of decoherence is now met~\cite{decoherence}.  Let
us stress how 
naturally this phenomenon appears: one can hardly talk about a
microsystem without taking it into account. The problem of studying
the dynamics of a microsystem in the present context coincides with
the well-known general problem of deriving master equations. It
appears here that the investigation of structures of the kind ${\hat
  a}_{\psi_0}^{\scriptscriptstyle\dagger} (t){\hat a}_{\psi_0}(t)$,
${\psi_0}$ involving the normal modes of a macrosystem is a very
natural starting point to derive master equations~\cite{me}. As a
consequence of interactions other modes of the system are involved.
One expects that at least for $t-t_0$ short enough a mixture
of states ${\psi^{\alpha}_t}$ replaces the pure state ${\psi_t}$,
the one-particle system associated to it being no longer isolated,
such that
\begin{displaymath}
   \textrm{Tr}\left[
{\hat a}_{\psi^{\alpha}_t}^{\scriptscriptstyle\dagger}
{\hat a}_{\psi^{\alpha}_t}
{\hat \varrho}^{\scriptscriptstyle (1)}_{t}\right] \ll 1;
\end{displaymath}
then the concept makes sense of binary ``collisions'' between these
one-particle states ${\psi_t}$ and the feeded normal modes of
the system. This situation is formally the same as the case of a
particle undergoing Brownian motion inside a macroscopic system;
models for such a case have been recently discussed inside many-body
scattering theory~\cite{qbm}. Extending slightly such a situation one
expects that the expression $\sum_{\alpha} \lambda^\alpha (t) {\hat
  a}_{\psi^{\alpha}_t}^{\scriptscriptstyle\dagger} {\hat
  a}_{\psi^{\alpha}_t}$ is 
obtained from ${\hat a}_{\psi_{0} }^{\scriptscriptstyle\dagger} {\hat
  a}_{\psi_{0} }$ by a linear map with a generalized Lindblad
structure. Then decoherence is made explicit and related to the
properties of the macroscopic system: this non Hamiltonian dynamics
already spoils the universal features of the dynamics of the
microsystem. At sufficiently longer times, as an increasing set of
states ${\psi^{\alpha}_t}$ is involved, there is no more a distinction
between the one-particle state and the other modes of the system.
Before this happens however the quantum behavior of the microsystem
had an essential role inside the dynamics of the macroscopic system. 

A final comment is in order now: we have seen that one reaches aspects
of quantum mechanics of a microsystem starting from a given
macrosystem. However in this way only a very particular sector of the
physics of the microsystem is used and explored; i.e., a particular
state $\psi_{0}$ is prepared, a not too long time evolution of it is
allowed, only some particular macroscopic variables can be effected by
the microsystem. The one-particle Hilbert space which comes into
evidence refers to the particular confinement region $\omega_1$. The
point is that this treatment encompasses all possible macrosystems
fitting in our model. Then the picture of a microsystem arises taking
into account all possible settings. Finally it is an idealization that
must be compatible with all of them: it is only at this stage that the
usual connection of microsystem with symmetry properties makes sense.
Endorsing this point of view microsystems are at the root of any
local process underlying the physics of all macrosystems; the
classical idea that these are structures composed of microsystems can
be taken on in quantum mechanics as a meaningful model only if
localization of microsystems in phase-space need not be better than
inside regions of value $\gg \hbar^3$: then as it is well-known a
phase-space density of particles, building the macrosystem, can be
introduced by means of a Wigner function related to ${\hat \varrho}_{t}$.

\section*{Acknowledgments}
L.~Lanz and B.~Vacchini gratefully acknowledge financial support by MURST
under Cofinanziamento. B.~Vacchini also acknowledges support by MURST under
Progetto Giovani.

\end{document}